\documentstyle[12pt]{article}
\newcommand{\postscript}[2]{\vspace{5cm}}

\parskip=\smallskipamount
\newcommand{\msb}{\mbox{$\overline{\rm{MS}}\ $}}
\newcommand{\mt}{\mbox{$m_t$}}
\newcommand{\mh}{\mbox{$M_H$}}
\newcommand{\mz}{\mbox{$M_Z$}}
\newcommand{\mw}{\mbox{$M_W$}}
\newcommand{\alsz}{\mbox{$\alpha_s(M_Z)$}}

\newcommand{\skipblk}[1]{}
\def\bqa{\begin{eqnarray}}
\def\eqa{\end{eqnarray}}

\newcommand{\ZP}[3]{Zeit. Phys. {\bf #1}, #2 (19#3)}

\newcommand{\PR}[3]{Phys. Rev. {\bf #1}, #2 (19#3)}
\newcommand{\PL}[3]{Phys. Lett. {\bf #1}, #2 (19#3)}
\newcommand{\NP}[3]{Nucl. Phys. {\bf #1}, #2 (19#3)}
\newcommand{\PRL}[3]{Phys. Rev. Lett. {\bf #1}, #2 (19#3)}

\newcommand{\con}[3]{ {\bf #1}, #2 (19#3)}

\newcommand{\etal}{{\em et al., }}

\newcommand{\ie}{{\em i.e., }}

\newcommand{\sinn}{\mbox{$\sin^2\theta_W\,$}}

\newcommand{\snu}{\mbox{$\stackrel{(-)}{\nu}$}}
\newcommand{\beq}{\begin{equation}}
\newcommand{\eeq}{\end{equation}}

\newcommand{\siz}{\mbox{$\sin^2\hat{\theta}_W(M_Z)\ $}}
\newcommand{\RA}{\mbox{$\rightarrow$}}

\def\mxth{\mathsurround=0pt }
\def\xversim#1#2{\lower2.pt\vbox{\baselineskip0pt \lineskip-.5pt
  \ialign{$\mxth#1\hfil##\hfil$\crcr#2\crcr\sim\crcr}}}

\def\ppm{\mbox{$\pm$}}
\def\beqa{\begin{eqnarray}}
\def\eeqa{\end{eqnarray}}
\def\ra{\rightarrow}
\title{Precision Tests of the Standard Model\thanks{Lectures
presented at {\it TASI-92}, Boulder, June 1992.}}
\author{Paul Langacker\\{\it Department of
 Physics, University of Pennsylvania}\\ {\it Philadelphia, PA 19104}\\
 \today,  UPR-0555T}
 \date{}
\begin{document}
\maketitle
\begin{abstract}
The implications of recent precision $Z$-pole, $W$ mass, and weak neutral
current data for testing the standard electroweak model, constraining the
$t$ quark and Higgs masses, \alsz,
and grand unification
are discussed. A fit to all data yields $\siz = 0.2328 \pm
0.0007$ (\msb) or $\sinn \equiv 1 - \mw^2/\mz^2 = 0.2267 \pm 0.0024$
(on-shell),
where the uncertainties are mainly from \mt. In the standard model one predicts
$\mt = 150^{+19 + 15}_{-24 - 20}$ GeV, where the central value assumes
\mh = 300 GeV and the second uncertainty is for \mh $\ra$ 60 GeV ($-$) or
1 TeV (+). In the minimal supersymmetric extension of the standard model
(MSSM) one predicts $\mt = 134^{+23}_{-28} \pm 5$ GeV, where the difference
is due the light Higgs scalar expected in the MSSM. There is no significant
constraint on \mh \ until \mt \ is known independently.
\end{abstract}

\section{Experimental Results}

Recent high precision measurements of  $Z$ pole observables by the
ALEPH, DELPHI, L3, and OPAL \cite{v1,4lep,ds} collaborations
at LEP and SLD at the SLC \cite{I2A},
the $W$ mass by
CDF \cite{v2} and UA2 \cite{v3}, atomic parity violation in cesium
\cite{v4,v5}, neutrino-electron scattering by CHARM II \cite{i1},
and other weak
neutral current observables \cite{iccfr,v6}, as well as the direct lower bounds
$m_t > 91$ GeV  (CDF \cite{v7}) and $M_H > 60$ GeV  (LEP average \cite{v8})
and the determination $\alsz = 0.12 \pm 0.01$ from $Z$-pole and
low-energy observables \cite{v9} allow precise tests of
the standard electroweak model and searches for certain types of new physics.
In
this talk, which is an update of previously presented analyses
\cite{ival}-\cite{ipol}, I review the status of the standard model tests
and parameters, and
the coupling constant predictions in ordinary and supersymmetric
grand unified theories (GUTs).

Many of the recent results
are summarized in Table \ref{data}. The LEP results are averages by D.
Schaile of the four LEP experiments as of March, 1993 \cite{ds}, which
includes nearly final results for the 1991 LEP run and contains a  proper
treatment of common systematic errors \cite{4lep}.
$M_Z$ is now known to the incredible precision of better than 0.01\%. This was
achieved by the method of resonant depolarization, in which the
(calculable) energies at which the small ($\sim 10 \%$) transverse polarization
of the leptons is destroyed by an oscillating $B$ field was used to
calibrate the energy of the LEP beams. The method is so precise that the tidal
effects of the moon, which cause the size of the LEP ring to change by a few
parts
in $10^8$ and thus change the energy by $\sim 8\ MeV$, had to be measured and
corrected for\footnote{This is the first experiment in which all four
interactions were important simultaneously!}.
 \begin{table}
\centering
\begin{tabular}{|l|l|l|} \hline
Quantity & Value & standard model \\ \hline
$M_Z$ (GeV) & $91.187 \pm 0.007$ & input   \\
$\Gamma _Z$ (GeV) & $2.491 \pm 0.007$ & $2.490 \pm 0.001 \pm 0.005
                                      \pm [0.006]$\\
$ R = \Gamma_{had}/ \Gamma_{l \bar{l}}$ & $20.87 \pm 0.07$ & $20.78
\pm  0.01 \pm 0.01 \pm [0.07]$ \\
$\sigma^h_p (nb)$ & $41.33 \pm 0.18$ & $41.42 \pm 0.01 \pm 0.01 \pm
                                                [0.06]$\\
$\Gamma_{b \bar{b}}$ (MeV) & $373 \pm 9$ & $375.9 \pm 0.2 \pm 0.5 \pm [1.3]$ \\
$A_{FB} (\mu)$ & $0.0152\pm 0.0027$ & $0.0141 \pm 0.0005 \pm 0.0010$
\\
$A_{pol} (\tau)$ & $0.140\pm 0.018$ & $0.137 \pm 0.002 \pm 0.005$ \\
$A_{e} (P_\tau)$ & $0.134\pm 0.030$ & $0.137 \pm 0.002 \pm 0.005$ \\
$A_{FB} (b)$ & $0.093\pm 0.012$ & $0.096 \pm 0.002 \pm 0.003$ \\
$A_{FB} (c)$ & $0.072\pm 0.027$ & $0.068 \pm 0.001 \pm 0.003$ \\
$A_{LR} $ & $0.100\pm 0.044$ & $0.137 \pm 0.002 \pm 0.005$ \\
                                                              \hline
$\Gamma_{l \bar{l}}$ (MeV) & $83.43 \pm 0.29$ & $83.66 \pm 0.02 \pm 0.13$ \\
$\Gamma_{had}$ (MeV) & $1741.2 \pm 6.6$ & $1739 \pm 1 \pm 4 \pm [6]$ \\
$\Gamma_{inv}$ (MeV) & $499.5 \pm 5.6$ &
$500.4 \pm 0.1 \pm  0.9$ \\
$N_{\nu}$ & $3.004 \pm 0.035 $ & $3$ \\

$\bar{g}_A$ & $-0.4999 \pm 0.0009$ & $-0.5$\\
$\bar{g}_V$ & $-0.0351 \pm 0.0025$ & $-0.0344\pm 0.0006 \pm 0.0013$
\\
$\bar{s}^2_W \ (A_{FB}(q))$& $0.2329\pm 0.0031$ & $0.2328 \pm 0.0003
\pm 0.0007 \pm$ ? \\
  \hline
$M_W$ (GeV) & $79.91 \pm 0.39$ & $80.18 \pm 0.02 \pm 0.13$ \\
$M_W/M_Z$ & $0.8813 \pm 0.0041$ & $0.8793 \pm 0.0002 \pm 0.0014$ \\
$Q_W (Cs)$ & $-71.04 \pm 1.58 \pm [0.88]$ & $-73.20 \pm 0.07
                                                    \pm 0.02$ \\
$g_A^e (\nu e \RA \nu e)$ & $-0.503 \pm 0.017$ & $-0.505 \pm 0 \pm 0.001$ \\
$g_V^e (\nu e \RA \nu e)$ & $-0.025 \pm 0.020$ & $-0.036 \pm 0.001
 \pm 0.001$ \\
\sinn & $ 0.2242 \pm 0.0042 \pm [0.0047]$ & $0.2269 \pm 0.0003
                                            \pm 0.0025$ \\ \hline
\end{tabular}
\caption[]{Experimental values for LEP \cite{v1,4lep,ds}
and SLC \cite{I2A} observables,
$M_W$ \cite{v2}, $M_W / M_Z$
\cite{v3}, the weak charge in cesium $Q_W$ \cite{v4,v5},
the parameters $g_{V,A}^e$ relevant to $\nu_\mu e$ scattering from CHARM II
\cite{i1}, and $\sinn \equiv 1 - M_W^2/M_Z^2$ from CCFR \cite{iccfr},
compared with the standard model predictions
for $M_Z = 91.187 \pm 0.007$ GeV, $m_t = 150 ^{+19}_{-24}$ GeV,
and 60 GeV $< M_H < 1$ TeV.  Only the first eleven $Z$-pole observables are
independent. The ? for the $\bar{s}^2_W \ (A_{FB}(q))$
prediction refers to the scheme dependence.
The two errors for $Q_W (Cs)$ and \sinn are experimental and
theoretical (in brackets).  The first error in the predictions is
from the uncertainties in $M_Z$ and $\Delta r$, the second is from
$m_t$ and $M_H$, and the third (in brackets) is the theoretical
QCD uncertainty for $\alsz = 0.12 \pm 0.01$  \cite{v9} .  The older
neutral current quantities described
in \cite{v6} are also used in the analysis.}
\label{data}
\end{table}
$\Gamma_Z, \Gamma_{l \bar{l}}, \Gamma_{had}$, $\Gamma_{b \bar{b}}$,
and
$\Gamma_{inv}$ refer respectively to the total, leptonic (average of $e, \mu,
\tau$), hadronic, $b \bar{b}$,
and invisible $Z$ widths; $ R \equiv \Gamma_{had}/
\Gamma_{l \bar{l}}$; $\sigma^h_p
= 12 \pi \Gamma _{e \bar{e}} \Gamma _{had} / M ^2_Z \Gamma^2_Z$ is the
hadronic cross section on the pole; and $N_\nu \equiv \Gamma _{inv} /
\Gamma _{\nu \bar{\nu}}$ is the number of light neutrino flavors.
A number of asymmetries have also been measured.
$A_{FB} (f)$ is the forward-backward asymmetry for $e^+ e^- \RA f \bar{f}$;
$A_{pol} (\tau)$ is the polarization of a
final $\tau$ ($L$ is positive), while $A_{e} (P_\tau)$ is essentially the
forward-backward asymmetry in the polarization; $A_{LR}$ is the
polarization asymmetry, which has recently been measured for
the first time by the SLD collaboration at the SLC \cite{I2A}.
 All of the asymmetries are Born contributions, from which various
QED, QCD, interference, and box contributions have been removed by
the experimenters.
Finally, $\bar{g}_A, \bar{g}_V$ are
effective Born couplings,  related, for example,
to $\Gamma_{l \bar{l}}$ and $A_{FB} (\mu)$
by\footnote{I assume lepton universality, throughout. This is
strongly supported by the LEP data.}
\beq
\Gamma_{l \bar{l}} = \frac{G_F M^3_Z}{6\sqrt{2} \pi} (\bar{g}^2_A
                                 + \bar{g}^2_V) \ \ \ \ \ \ \
A_{FB}(\mu) = \frac{3 \  \bar{g}^2_V \  \bar{g}^2_A}
      {(\bar{g}^2_V + \bar{g} ^2 _A)^2}.
\eeq
Similarly,
\beq A_{LR} = \frac{2 \  \bar{g}_V \  \bar{g}_A}
      {\bar{g}^2_V + \bar{g} ^2 _A},  \eeq
with the same expression for
$A_{pol} (\tau)$ and $A_{e} (P_\tau)$.

Of the $Z$-pole  observables only $M_Z$, $\Gamma_Z, R,$ $\sigma^h_p$,
 $\Gamma_{b \bar{b}}$, $A_{FB} (\mu)$,  $A_{pol} (\tau)$, $A_{e} (P_\tau)$,
$A_{FB}(b)$ (which is corrected for $b\bar{b}$ oscillations),
$A_{FB}(c),$ and $A_{LR}$
are used in the analysis. $ \Gamma_Z,\ R,$ and $\sigma^h_p$ are used
rather than the more physically transparent $ \Gamma_Z,\ \Gamma_{l \bar{l}},$
and $ \Gamma_{had}$ because the former are closer to what is actually
measured and are relatively weakly correlated. (The combined LEP
values \cite{ds} for the correlations are used.)
$\bar{s}^2_W \ (A_{FB}(q))$,
which is the effective weak angle obtained from the
charge asymmetry in hadronic decays, is not used because the results have
only been presented assuming the validity of the standard model. The other
LEP observables are not independent but are displayed
for completeness.

Recent measurements of the $W$ mass and weak neutral
current data are also displayed in Table \ref{data}.  $Q_W (Cs)$ is the
effective charge of the parity-violating interaction in cesium \cite{v4},
while $g_{V,A}^e$ are the coefficients of the vector and axial electron
currents in the effective four-fermi interaction for $\snu_\mu e
\RA \snu_\mu e$ as obtained by  CHARM II \cite{i1}.
The preliminary value of the on-shell weak angle
$\sinn \equiv 1 - M_W^2/M_Z^2 = 0.2242 \pm 0.0042 \pm [0.0047]$
obtained from deep inelastic neutrino scattering
from CCFR \cite{iccfr} at Fermilab is in reasonable
agreement with the earlier
CERN values $0.228 \pm 0.005 \pm [0.005]$ \cite{cdhs}, and
$0.236 \pm 0.005 \pm [0.005]$ \cite{charm}, though the central
value is somewhat lower. The errors in brackets are theoretical.
They are dominated by the $c$-quark threshold in the charged current
scattering
used to normalize the neutral current process, and are
strongly correlated between the experiments. Older neutral current
results, included in the analysis, are described in~\cite{v6}.

The
standard model predictions for each quantity other than $M_Z$ are also
shown.
These are computed using $M_Z = 91.187 \pm 0.007$ GeV
as input, using the range of $m_t$ determined from the global fit and
60 GeV $< M_H < 1$ TeV. The agreement is excellent.

The $b$ observables $\Gamma_{b\bar{b}}$ and $ A_{FB} (b)$ are especially
important because the predictions depend on
the $SU_2$ assignments of the $b$.  In Table~\ref{b} the
experimental values are compared with topless models and
other
alternatives with $V+A$ currents.  It is seen that the data uniquely
picks out the standard model from these alternatives~\cite{v15}. This
conclusion is strenghtened by a recent detailed analysis by Schaile and
Zerwas \cite{v16} of LEP and lower energy data, which  yields
\beq  t_{3L}(b) = -0.490^{+0.015}_{-0.012}\ \ \ \ \ \ \ \
t_{3R}(b) = -0.028 \pm 0.056 \eeq
for the third component of the weak isospin of the $b_{L,R}$, respectively,
in agreement with the standard model expectations of $-1/2$ and $0$ --
\ie topless models are excluded and the $b_L$ must be
 in a weak doublet with
the $t_L$.
\begin{table}                   \centering
\small \def\baselinestretch{1} \normalsize
\begin{tabular}{|c|c|c|c|c|c|} \hline
Quantity & Experiment & SM & Topless & Mirror & Vector \\
\hline
$\Gamma_{b\bar{b}}$ (MeV)&
373 \ppm 9 & 376 & 24 & 376 & 728 \\
$A_{FB} {(b)}$ & 0.093 \ppm 0.012 & 0.096 & 0 & $-0.096$ & 0 \\ \hline
\end{tabular}
\caption[]{Predictions of the standard model (SM),
topless models, a mirror
model with $(t \; b)_R$ in a doublet, and a vector model with left and
right-handed doublets, for $\Gamma_{b\bar{b}}$
and $A_{FB} (b)$, compared with the experimental values.}
\label{b}
\end{table}

\section{Standard Model Tests and $m_t$}
Results will be presented in the $\overline{\rm{MS}}$ \cite{v17} and
on-shell \cite{v18} schemes. I use the radiative corrections calculated by
Degrassi {\it et al.}
\cite{v19} for the $W$ and $Z$ masses, those of Hollik
\cite{v20} for the $Z$ widths, and  generalized Born expressions for
the Born contributions to the asymmetries. The latter are obtained
from the data, {\it e.g., } by using the program
ZFITTER \cite{v22}. The calculations in \cite{v19}-\cite{v22} are in
excellent agreement with each other and with those in \cite{v21}.
Radiative corrections to low energy
neutral current processes are described in \cite{v6}.

In the standard model
\begin{eqnarray}
M^2_Z & = & \frac{A^2_0}
{\hat{\rho} \hat{c}^2 \hat{s}^2 (1 - \Delta \hat{r}_W)}
=  \frac{A^2_0}{c^2 s^2 (1-\Delta r)} \nonumber \\
M^2_W & = & \hat{\rho} \hat{c}^2 M^2_Z =  c^2 M^2_Z  \label{mwz}
\end{eqnarray}
where $A^2_0 =  \pi \alpha/ \sqrt{2} G_F = (37.2803 \  \rm{GeV})^2,
\hat{s}^2 \equiv$ \siz  refers to the weak angle in the
$\overline{\rm{MS}}$
scheme \cite{v17}, $s^2 \equiv$ \sinn $= 1 - M^2_W / M^2_Z$ refers to
the on-shell scheme \cite{v18}, $\hat{c}^2 \equiv 1 - \hat{s}^2$, and
$c^2 \equiv 1 -s^2$;  $\Delta
\hat{r} _W, \hat{\rho} - 1$, and $\Delta r$ are radiative correction
parameters.  As is well known \cite{v23}, $\hat{\rho}
\sim 1 + \Delta \rho_t$, where
\begin{equation}
\Delta \rho_t = \frac{3G_F m_t^2}{8 \sqrt{2} \pi^2} \simeq 0.0031
\left( \frac{m_t}{100 \rm{GeV}}\right)^2 \ , \label{rhat}
\end{equation}
has a strong $m_t$ dependence, while
$\Delta r \simeq \Delta r_0 - \Delta \rho_t / \tan^2 \theta_W$ is
even more sensitive.
$\Delta \hat{r}_W \sim \Delta r_0 \sim  1 - \alpha/\alpha(M_Z) \sim
0.07$ has no quadratic
$m_t$ dependence.  There is additional logarithmic dependence on $m_t$
and $M_H$ in $\hat{\rho}, \Delta \hat{r}_W$, and $\Delta r$, as
well as $O (\alpha)$ effects associated with low energy physics.  These
effects are important and are fully incorporated in the analysis, but
will not be displayed here.

 Gluonic corrections
to $ \Delta \rho_t$ of order $-\alpha \alpha_s m_t^2/M_Z^2$  can be
important for large $m_t$ \cite{v24,v25}. The leading perturbative term is
\cite{v24} $-2
\alpha_s(m_t) (\pi^2 + 3)/9 \pi \sim -0.10$ times the expression in
(\ref{rhat}). These corrections, which increase the
predicted value of $m_t$ by about 5\%, are included in
the analysis\footnote{These
terms were omitted from previous analyses (e.g., \cite{ival}), because of
 uncertainties in both the magnitude and sign of important nonperturbative
effects \cite{v25}. However, a careful new analysis \cite{v25a} indicates
that the perturbative esimate is an excellent approximation. A future
global analysis will include small additional corrections.}.

The most accurate determination of \siz and \sinn are from $M_Z =
91.187 \pm 0.007$~GeV.
The values are shown in
Table~\ref{tmz}
for $m_t = 100$ and 200~GeV and $M_H = 300$~GeV, and also for the
global best fit range for $m_t$ and $M_H$.  It is apparent that
the extracted value of $\sin^2\theta_W$ depends strongly on $m_t$,
while $\sin^2\hat{\theta}_W (M_Z)$ is considerably less
sensitive due to the smaller coefficient of the quadratic $m_t$ term in $
\hat{\rho}$ than in $\Delta r$.
For fixed $m_t$
and $M_H$ the
uncertainty of $\pm 0.0003$ in $\sin^2\theta_W$ has two components:
the experimental error from $\Delta M_Z$ is only $\pm 0.0001$, while
the theoretical error (from the uncertainty of $\pm 0.0009$ in $\Delta
r$ from low energy hadronic contributions) is larger, $\pm 0.0003$.
The $\pm 1 \sigma$ limits on $\sin^2 \hat{\theta}_W(M_Z)$ as a
function of $m_t$ are shown in Figure~\ref{mtop}.

\begin{table}\centering
\small \def\baselinestretch{1} \normalsize
\begin{tabular}{|ccc|} \hline
$m_t$ (GeV)  & $\sin^2\theta_W$    & $\sin^2\hat{\theta}_W(M_Z)$ \\
\hline
100      & $0.2322 \pm 0.0003$    & $ 0.2340 \pm 0.0003$ \\
200      & $0.2204 \pm 0.0003$    & $0.2311 \pm 0.0003$ \\
$150^{+19}_{-24}$ & $0.2269 \pm 0.0025$ & $0.2328 \pm 0.0007$ \\
\hline
\end{tabular}
\caption[]{Values of $\sin^2\theta_W = 1 - M^2_W/M^2_Z$ and
$\sin^2\hat{\theta}_W(M_Z)$ obtained from $M_Z = 91.187 \pm
0.007$~GeV, assuming $(m_t, M_H) =$~(100, 300) and (200, 300) GeV.  In
the last row $m_t = 150^{+19}_{-24}$~GeV (obtained from the global fit
to all data) and 60~GeV$< M_H < 1000$~GeV.}
\label{tmz}
\end{table}

\begin{figure}
\small \def\baselinestretch{1} \normalsize
\postscript{xxmt.ps}{0.8}
\caption[]{One $\sigma$ uncertainty in $\sin^2\hat{\theta}_W(M_Z)$ as a
function of $m_t$ for $\sin^2\hat{\theta}_W(M_Z)$ determined from
various inputs for $M_H = 300$~GeV.  The direct lower limit $m_t >
91$~GeV and the 90\% CL fit to all data are also shown.}
\label{mtop}
\end{figure}

The ratio $M_W/M_Z = 0.8813 \pm 0.0041$ determined by UA2 \cite{v3}
and $M_W =
79.91 \pm 0.39$~GeV from CDF \cite{v2} determine the values of $\sin^2
\hat{\theta}_W (M_Z)$ shown in Table~\ref{stab}.  From
Figure~\ref{mtop}                           it is apparent that
$M_Z$, $M_W$, and $M_W/M_Z$ together imply an upper limit of
$O$(200~GeV) on $m_t$.  A simultaneous fit of $M_Z$, $M_W$, and
$M_W/M_Z$ to $\sin^2\hat{\theta}_W(M_Z)$,        and $m_t$ yields $m_t
= 145^{+42}_{-49} \pm 16$~GeV, where the second uncertainty is from $M_H$ in
the range 60-1000~GeV, with a central value of 300~GeV.  The
90(95)\%~CL upper limits on $m_t$ are 211~(223)~GeV
(Table~\ref{ttab}).  The upper limits are for $M_H = 1000$~GeV,
which gives the weakest constraint.  The value of
$\sin^2\hat{\theta}_W(M_Z)$, including the uncertainties from $m_t$ and
$M_H$, is also given in Table~\ref{ttab}.

\begin{table}\centering  \small \def\baselinestretch{1} \normalsize
\begin{tabular}{|c|c|c|c|} \hline
Data & $\sin^2\hat{\theta}_W(M_Z)$  & \ \      & \ \   \\
\ \  &   $m_t = 100$      & $m_t=200$& $m_t=150^{+19}_{-24}$ \\ \hline
$M_Z$ & 0.2340 \ppm 0.0003 & 0.2311 \ppm 0.0003 & 0.2328 \ppm 0.0007
\\
$M_W, \frac{M_W}{M_Z}$ & 0.2331 \ppm 0.0022 & 0.2345 \ppm 0.0022 &
0.2339 \ppm 0.0022 \\
$ \Gamma_{Z},R,\sigma^h_p $ & 0.2333 \ppm 0.0006 &
0.2319 \ppm 0.0006 & 0.2327 \ppm
0.0006 \\
$\Gamma_Z$ & 0.2332 \ppm 0.0006 & 0.2320 \ppm 0.0006 & 0.2327 \ppm
0.0007 \\
$\Gamma_{l\bar{l}}$ & 0.2340 \ppm 0.0007 & 0.2326 \ppm 0.0007 & 0.2333
\ppm
0.0008 \\
$\Gamma_{b\bar{b}}$ & 0.236 \ppm 0.004 & 0.232 \ppm 0.004 & 0.234
\ppm
0.004 \\
$\Gamma_Z/M_Z$ & 0.2329 \ppm 0.0009 & 0.2323 \ppm 0.0009 & 0.2326
\ppm
0.0009 \\
$\Gamma_{l\bar{l}}/M_Z$ & 0.2341 \ppm 0.0010 & 0.2332 \ppm 0.0010 & 0.2336
\ppm 0.0010 \\
$\Gamma_Z/M^3_Z$ & 0.230 \ppm 0.003 & 0.236 \ppm 0.003 & 0.232 \ppm
0.004 \\
$\Gamma_{l\bar{l}}/M_Z^3,R$ & 0.231 \ppm 0.004 & 0.234 \ppm 0.003 & 0.232
\ppm 0.004 \\
$A_{FB} (\mu)$ & 0.232 \ppm 0.002 & 0.232 \ppm 0.002 & 0.232
\ppm
0.002 \\
$A_{pol} (\tau)$ & 0.232 \ppm 0.002 & 0.232 \ppm 0.002 & 0.232 \ppm
0.002 \\
$A_{FB} (b)$ & 0.233 \ppm 0.002 & 0.233 \ppm 0.002 & 0.233 \ppm
0.002 \\
$A_{LR}$ & 0.238 \ppm 0.006 & 0.238 \ppm 0.006 & 0.238 \ppm
0.006 \\
$\nu N \ra \nu X$ & 0.233 \ppm 0.005 & 0.238 \ppm 0.005 & 0.235 \ppm
0.005 \\
$\nu p \ra \nu p$ & 0.212 \ppm 0.032 & 0.212 \ppm 0.031 & 0.212 \ppm
0.032 \\
$\nu_\mu e \ra \nu_\mu e$ & 0.232 \ppm 0.009 & 0.231 \ppm 0.009 &
0.232 \ppm 0.009 \\
$e^{\uparrow \!\downarrow} D \ra eX$ & 0.222 \ppm 0.018 & 0.223 \ppm
0.018 & 0.222 \ppm 0.018 \\
atomic parity & 0.224 \ppm 0.008 & 0.221 \ppm 0.008 & 0.223 \ppm 0.008
\\
All & 0.2337 \ppm 0.0003 & 0.2314 \ppm 0.0003 & 0.2328 \ppm 0.0007 \\
\hline
\end{tabular}
\caption[]{Values of $\sin^2\hat{\theta}_W (M_Z)$ obtained from various
inputs, for $(m_t, \; M_H) =$~(100, 300) and (200, 300)~GeV.  In the
last column $m_t = 150^{+19 +15}_{-24 -20}$~GeV, from the global fit,
correlated with
60~GeV $< M_H < 1000$~GeV.  For $\nu N \ra \nu X$ the uncertainty includes
0.003
(experiment) and 0.005 (theory). For atomic parity, the experimental and
theoretical components of the error are 0.007 and 0.004 respectively.}
\label{stab}
\end{table}

\begin{table}    \centering
\small \def\baselinestretch{1} \normalsize
\begin{tabular}{|l|l|l|l|} \hline
data & \siz & $m_t$ (GeV) & $m^{max}_t$ (GeV) \\ \hline
$M_Z, M_W,M_W/M_Z$ & $0.2329 \pm 0.0014$ & $145^{+42}_{-49} \pm 16$ &
$211 (223)$\\
$Z$-POLE & $0.2328 \pm 0.0008$ &
$150 \pm 27 \pm 18$   & $198 \  (206)$\\
$Z$-POLE $ + M_W, M_W/M_Z$ & $0.2328 \pm 0.0007$
                      & $150^{+21}_{-26}
                     \pm 18$ & $193   (200)$\\
$M_Z, \nu N$ & $0.2333 ^{+0.0011}_{-0.0016}$ &
$132^{+54}_{-40} \pm 19$
                      & $210 \  (223)$\\
All & $0.2328 \pm 0.0007$ & $150 ^{+19 +15}_{-24 -20}$ & $190 \  (197)$
                                     \\ \hline
\end{tabular}
\caption[]{
Values of \siz and $m_t$ obtained for various data sets.
$Z$-POLE refers to the first 11 constraints in Table \ref{data}
 (with correlations).
The \siz error includes $m_t$ and $M_H$. The first error for $m_t$
includes experimental and theoretical uncertainties for $M_H = $ 300 GeV.
The second error is the variation for $M_H \rightarrow 60$ GeV $(-)$ and
$M_H \rightarrow 1000$ GeV $(+)$. The last column lists the upper limits
on $m_t$ at 90 (95)\% CL for $M_H = 1000$ GeV, which gives the weakest
upper limit. The direct CDF constraint $m_t > 91$ GeV is included.}
\label{ttab}
\end{table}

The partial width for $\Gamma \ra f \bar{f}$ is given
by~\cite{v20,v21},
\beq \Gamma_{f\bar{f}} = \frac{C_f \hat{\rho} G_F M_Z^3}{6
\sqrt{2} \pi} \left( |a_f|^2 + |v_f|^2 \right). \label{eq27} \eeq
  The axial and vector couplings are
\beqa a_f &=& t_{3L} (f) = \pm \frac{1}{2} \nonumber \\
v_f &=& t_{3L} (f) - 2  \sin^2\hat{\theta}_W (M_Z) q_f, \label{eq28}
\eeqa
where $t_{3L}(f)$ and $q_f$ are respectively the third component
of weak isospin and electric charge of fermion $f$;
$\hat{\rho}$ is                             dominated by the
$m_t$ term (cf (\ref{rhat}).  The coefficient     comes about by
rewriting the tree-level formula
\beq \frac{g^2(M_Z) M_Z}{8 \cos^2 \theta_W} = \frac{G_F
M_Z^3}{\sqrt{2}}. \label{eq29} \eeq
Expressing the width in this way incorporates the bulk of the
radiative
corrections, except for the large $m_t$ dependence in $\hat{\rho}$.
Additional small radiative corrections are included but not displayed
here.
             The factor in front incorporates the color factor and QED
and QCD corrections:
\beq C_f = \left\{ \begin{array}{lll}
1 + \frac{3\alpha}{4\pi} q_f^2 & ,& {\rm leptons} \\
3 \left( 1 + \frac{3\alpha}{4\pi} q^2_f \right) \left( 1 +
\frac{\alpha_s}{\pi} + 1.405 \frac{\alpha_s^2}{\pi^2} \right)
\ &,& {\rm quarks} \end{array} \right.      \label{eq30} \eeq
where the range
 $\alpha_s (M_Z) \simeq 0.12 \pm 0.01$ from $Z$-decay event
topologies and other data \cite{v9} is used.

(\ref{eq27}) is written neglecting the fermion masses.  In practice,
fermion
mass corrections~\cite{v20} must be applied for $\Gamma_{b\bar{b}}$.
They are also included in the following for
$\Gamma_{c\bar{c}}$ and $\Gamma_{\tau \bar{\tau}}$, though
the effects are small. There are significant correlations between
the experimental values of the various total and partial $Z$ widths,
which must be included in a global analysis.

The vertex corrections for $\Gamma_{b\bar{b}}$
depend strongly on $m_t$ and must be included as an
extra correction~\cite{v26}.  For fixed $M_Z$ the $b\bar{b}$ width
actually decreases
with $m_t$, while the other modes all increase (because of the
$\hat{\rho}$ factor).  This gives a means of separating
 $\hat{\rho} (m_t)$ from such new physics as nonstandard Higgs
representations by comparing $\Gamma_{b\bar{b}}$ or
$\Gamma_Z$ with the other data~\cite{v10}.

The standard model predictions for $\Gamma_Z$, $\Gamma_{\ell^+\ell^-}
(\ell = e, \mu,$ or $\tau), \; R \equiv \Gamma_{had}
/\Gamma_{\ell^+\ell^-}$, and the invisible width $\Gamma_{inv}$
as a function of $m_t$ are compared with the experimental
results in Figures~\ref{gam1} and \ref{gam2}.  ($\sin^2
\hat{\theta}_W(M_Z)$ in $v_f$ is obtained from $M_Z$).  One sees that
the agreement is excellent for $m_t$ in the 100 -- 200 GeV range.  The
results of fits to the $Z$ widths are listed in Tables \ref{stab} and
\ref{ttab}.   The $R$ ratio, which is insensitive to $m_t$, is slightly above
the standard model prediction, though only at the 1$\sigma$ level.  As
will be discussed, $R$ favors a slightly higher value of \alsz than the
value obtained from event topologies and low energy data.

The invisible width in Figure~\ref{gam2} is
clearly in agreement with $N_\nu = 3$ but not $N_\nu = 4$. In fact, the
result \cite{ds} $N_\nu = 3.004 \pm 0.035$ not only eliminates extra
fermion families with $m_\nu \ll M_Z/2$, but also supersymmetric
models with light
sneutrinos ($\Delta N_\nu = 0.5$) and models with triplet ($\Delta N_\nu =
2$) or doublet ($\Delta N_\nu = 0.5$) Majorons \cite{numass}.
$N_\nu$ does not include
sterile ($SU_2$-singlet) neutrinos. However, the complementary bound
$N_\nu' < 3.3$ (95\% CL) from nucleosynthesis \cite{v27} {\it does}
include sterile neutrinos for a wide range of masses and mixings, provided
their mass is less than $\sim 20$ MeV.

\begin{figure}
\small \def\baselinestretch{1} \normalsize
\postscript{xxgamf.ps}{0.8}
\caption[]{Theoretical predictions for $\Gamma_Z,\Gamma_{\ell^+\ell^-}$,
and $R = \Gamma_{had}/\Gamma_{\ell^+\ell^-}$ in the standard model
as a function of $m_t$, compared with the experimental
results. The $M_H$ dependence is too small to see on the scale of the
graph.
The QCD uncertainties in $\Gamma_Z$ and
$R$ are indicated.}
\label{gam1}
\end{figure}

\begin{figure}
\small \def\baselinestretch{1} \normalsize
\postscript{xxginv.ps}{0.55}
\caption[]{Theoretical prediction for $\Gamma_{inv}$ in the standard
model with $N_\nu = 3$ and 4, compared with the experimental
value.}
\label{gam2}
\end{figure}

One can obtain  precise $(\Delta = O (\pm 0.0007))$ values of
$\sin^2\hat{\theta}_W(M_Z)$ from $\Gamma_Z$ and $\Gamma_{\ell^+
\ell^-}$
(Table~\ref{stab}).  The major sensitivity is through the $M^3_Z$
factor in (\ref{eq27}) rather than from the vertices (\ie  the $v_f$).
It is useful to also obtain the $\sin^2\hat{\theta}_W(M_Z)$ from the
vertices.
Values can be obtained from the ``reduced widths''
$\Gamma_Z/M^3_Z, \; \Gamma_{\ell^+\ell^-}/M_Z^3$, and $R$.  As can be
seen in Table~\ref{stab}, the $\sin^2\hat{\theta}_W(M_Z)$
sensitivity from
$\Gamma_Z/M_Z^3$ and the combination $(\Gamma_{\ell^+\ell^-}/M_Z^3,
R)$ is around $\pm 0.004\; (\Gamma_{\ell^+\ell^-}/M_Z^3$ and $R$
individually give large asymmetric errors).  Yet another determination
of $\sin^2\hat{\theta}_W(M_Z)$ comes from $\Gamma_Z/M_Z$ and
$\Gamma_{l^+l^-}/M_Z$.  As can be seen in Table~\ref{stab} the
values obtained are insensitive to $m_t$.  This can be understood from
(\ref{mwz}) and (\ref{eq27}), from which one sees that
$\Gamma_{f\bar{f}}/M_Z$ has no quadratic $m_t$ dependence (except
$f=b)$.  Of course, the various values of $\sin^2\hat{\theta}_W$
obtained from the $\Gamma$'s are not all independent.

At tree level the asymmetries can be written
\beq  A_{FB}(f)  \simeq 3 \eta_e \eta_f, \eeq
and \beq  A_{pol}(\tau) \simeq 2 \eta_\tau, \eeq
where
\beq  \eta_f \equiv \frac{v_f a_f}{v_f^2 + a_f^2}, \eeq
and $v_f$ and $a_f$ are the tree-level vector and axial couplings in
(\ref{eq28}).
These expressions are an excellent first approximation even in
the presence of higher-order corrections, provided that $v_f$ is expressed
in terms of \siz, {\it i.e.,} one identifies $v_f$ and $a_f$ with the
effective
Born couplings $\bar{g}_V$ and $\bar{g}_A$.
$ A_{FB}(b) = 0.093 \pm 0.012$ has been corrected for $b
\bar{b}$ oscillations \cite{v1}, using
\beq A_{FB}(b) = \frac{ A_{FB}^{obs}(f)}{1-2 \chi}, \eeq
where $\chi = 0.126 \pm 0.012$ is the oscillation probability at the
$Z$-pole. $Z b \bar{b}$ vertex corrections can be added to $ A_{FB}(b)$ but
are negligible numerically. The predictions for $A_{FB} (\mu)$,
$A_{pol} (\tau)$, and $A_{FB} (b)$ are compared with the experimental
data in Figure~\ref{asym}. Again, the agreement is excellent.

\begin{figure}
\small \def\baselinestretch{1} \normalsize
\postscript{xxafb.ps}{0.8}
\caption[]{Theoretical prediction for $A_{FB} (\mu)$, $A_{pol} (\tau)$, and
$A_{FB} (b)$ in the standard
model as a function of $m_t$ for $M_H$ = 60 (dotted line), 300 (solid), and
1000 (dashed) GeV, compared with the experimental values. The
theoretical uncertainties from $\Delta \Delta
r = \pm 0.0009$ are also indicated.}
\label{asym}
\end{figure}

The results for $\sin^2\hat{\theta}_W$ obtained from a variety of low
energy neutral current processes are listed in Table~\ref{stab}.
The values obtained from atomic parity violation, $e^{\uparrow
\downarrow}D$, and $\nu e$ and $\nu p$ elastic scattering are
consistent with the value obtained from $M_Z$.  They all have a
similar dependence on $m_t$ as the $M_Z$ value and therefore do not
significantly constrain $m_t$. They are, however, quite important in
searches for new physics.

On the other hand, the value of the on-shell $\sin^2\theta_W$ obtained
from deep inelastic $\nu N$ scattering
\cite{v28} is insensitive to $m_t$.
As can be seen in Table~\ref{stab} and
Figure~\ref{mtop} the corresponding $\sin^2\hat{\theta}_W (M_Z)$
increases rapidly with $m_t$.  From Table~\ref{ttab} deep inelastic
$\nu N$ scattering (combined with $M_Z$) gives $ m_t < 210(223)$~GeV
at 90(95)\% CL. These are somewhat weaker than previous limits (193 (207)
GeV) \cite{ival} due to the inclusion of the new CCFR result \cite{iccfr},
with its slightly lower value for \sinn (+6 GeV) and due to the
inclusion of $O(\alpha \alpha_s m_t^2)$ radiative corrections (+11 GeV).

The results of global fits to all data
are shown in
Tables~\ref{stab} and \ref{ttab}.  All results include full statistical and
systematic uncertainties in the experimental data as well as all of the
important correlations.
In particular, one
obtains the prediction\footnote{This is in excellent agreement with the result
$148^{+18 +17}_{-20 -19}$ of Schaile \cite{ds}.}
\beq m_t = 150^{+19 +15}_{-24 -20}  \ {\rm GeV}, \label{eq52} \eeq
where the central value assumes \mh \ = 300 GeV.
The second error is from the Higgs mass, assuming $60$~GeV
$<M_H< 1000$~GeV. The \mt \ and \mh \ dependences are strongly correlated.
The relation between the two in the radiative corrections is not universal,
but  a reasonable interpolation of the \mh dependence is
\beq m_t {\rm (GeV)} = 150^{+19}_{-24} + 12.5 \ln (\mh/300 {\rm GeV}).
\label{eq52a} \eeq
Alternately, we can allow \mh \ to be a free parameter in the
range 60 -1000 GeV, with the result that $m_t = 131^{+47}_{-28}$
GeV, with the lower central value occurring because the best fit
is for \mh \ = 60 GeV.

 The upper limit on \mt \ is
\beq m_t < \left\{ \begin{array}{clc} 190 \ {\rm GeV},
 & 90\%& \  CL \\
197\ {\rm GeV}, & 95\%& \ CL \\ 208\ {\rm GeV}, & 99\%& \ CL
\end{array}
\right.\ \ \ \ \ , \label{eq53} \eeq
which occurs for $M_H = 1000$~GeV.  For $M_H = 60 (300)$~GeV, the 90\%
CL limit is 158 (175)~GeV and the 95\%
CL limit is 165 (182)~GeV. The upper and lower limits on $m_t$ are
shown as a function of $M_H$ in Figure \ref{ht}.
The values of $\sin^2\hat{\theta}_W$ and
$m_t$ and the $m_t$ limits for various subsets of the data are given
in Table~\ref{ttab}.  The $\chi^2$ distribution as a function of
$m_t$ is shown in Figure~\ref{chis} for $M_H = $~60, \mz, 300, and
1000~GeV. The fit is excellent\footnote{In fact, the fit is {\em too}
good. This has always been the case for precision neutral current
and $Z$-pole experiments \cite{v6}. The most likely explanation is a tendency
for experimenters to overestimate systematic errors.}, with a $\chi^2/df$ of
168/206 $\sim$ 0.82 for $m_t = 150,\ M_H = 300$ GeV.

The result in (\ref{eq52}) is very close to the value
$149^{+21}_{-27} \pm 16 $ obtained about 1 year ago. The agreement
is somewhat fortuitous: the new 1991 LEP and other data lower
the prediction by $\sim 9$ GeV, but this is compensated by
the inclusion of $O(\alpha \alpha_s m_t^2)$ radiative corrections
(+8 GeV) and the use of 300 (rather than 250) GeV as the central
\mh \ value (+2 GeV).

\begin{figure}
\small \def\baselinestretch{1} \normalsize
\postscript{xxmtmh.ps}{0.55}
\caption[]{Best fit value for $m_t$ and upper and lower limits as a
function of $M_H$. The direct lower limit $M_H >$ 60 GeV \cite{v8} and
the approximate triviality limit \cite{v29} $M_H < $ 600 GeV are also
indicated. The latter becomes $M_H < $ 200 GeV if one requires that the
standard model holds up to the Planck scale.}
\label{ht}
\end{figure}
\begin{figure}
\small \def\baselinestretch{1} \normalsize
\postscript{xxchis.ps}{0.55}
\caption[]{$\chi^2$ distribution
for all data (207 df) in the standard model
 as a function of $m_t$, for $M_H = 60$, \mz, 300, and
1000~GeV. The direct constraint \mt $> 91$ GeV is {\em not}
included.} \label{chis}
\end{figure}

The prediction in (\ref{eq52}) is for the minimal standard model.
In the minimal supersymmetric extension (MSSM),
for almost all of the allowed parameter range for the superpartner
spectrum the only significant effect on the analysis is in the Higgs
sector \cite{susyrad}. There is a light ($M < 150$ GeV)  scalar which
acts like
a light standard model Higgs
(as far as radiative corrections are concerned), and
(typically)  the other Higgs particles and superpartners do not contribute
significantly. Thus, for the MSSM we will take 60 GeV $< \mh <$ 150 GeV with a
central  of \mz , yielding:
\beq
{\rm MSSM:} \ \ \ \ \
m_t = 134^{+23}_{-28} \pm 5  \ {\rm GeV}. \label{eq52b}
\eeq
For  \mh \ a free parameter in the
range 60 -150 GeV, one obtains $m_t = 131^{+31}_{-28}$
GeV, with the  best fit
for \mh \ = 60 GeV.

The data also yield an indirect {\it lower} limit on $m_t$ (Figure
\ref{mtop}). For $M_H$ = 60 GeV
one obtains $m_t > $ 95(83) GeV at 90(95)\% CL.
The corresponding limits are 118(108) GeV for $M_H = 300$ GeV and
138(129) GeV for $M_H = 1000$ GeV. The lower bound is comparable to the
direct CDF limit $m_t > 91$ GeV (95\% CL) \cite{v7}. However, it is more
general in that it applies even for nonstandard $t$ decay modes, for which
the direct lower limit is $\sim 60$ GeV.

The data will not significantly constrain $M_H$ until $m_t$ is
known separately. At present the best fit occurs
for lower values of
$M_H$, but the change in $\chi^2$ between $M_H$ = 60 and 1000 GeV is
only 0.6. From Figure \ref{chis} is it obvious that if $m_t$ is measured
directly to within 5-10 GeV it may be possible to constrain $M_H$,
particularly if $m_t$ is in the lower part of the allowed range. This
is further illustrated in Figure \ref{mtdirect}, in which are displayed
the 68 and 90\% CL \mh \ ranges that could be obtained from present data
if \mt \ were known to 10 GeV.
\begin{figure}
\small \def\baselinestretch{1} \normalsize
\postscript{xxhiggs.ps}{0.55}
\caption[]{68 and 90\% CL \mh \ ranges that could be obtained from present data
if \mt \ were known by direct measurement to $\pm$ 10 GeV as a function of the
central value of \mt.} \label{mtdirect} \end{figure}

Assuming the standard model, one therefore concludes 91~GeV~$< m_t <
197$~GeV at 95\% CL.  In
most cases, the effect of new physics is to
{\em strengthen} the upper bound rather than weaken it.  The obvious
question is, why is $m_t$ so large (or why are the other fermion
masses so small)?
Note that the value of \mt \ considered here is the position of
the pole in the $t$ propagator (not the running mass). It should
coincide (with a theoretical ambiguity of
$\sim $ 5 GeV) with the kinematic mass
relevant for the production of the $t$ quark at hadron colliders.

For the weak angle one obtains (in the standard model)
\beqa \sin^2\theta_W & = & 0.2267 \pm 0.0024 \nonumber \\
\sin^2\hat{\theta}_W (M_Z) & = & 0.2328\pm 0.0007, \label{eq54} \eeqa
where the uncertainty  is mainly from $m_t$.
The small uncertainty from $M_H$ in the range 60 - 1000~GeV is
included in the errors in (\ref{eq54}).
The corresponding value in the MSSM is
$\sin^2\hat{\theta}_W (M_Z)
=  0.2326\pm 0.0006$.
 Of course, $\sin^2
\hat{\theta}_W (M_Z)$ is much less sensitive to $m_t$ and $M_H$ than
$\sin^2\theta_W$.  All of the values obtained from individual observables
are in excellent agreement with (\ref{eq54}).  In particular, the
$\sin^2\hat{\theta}_W$ values obtained assuming $m_t =
150^{+19}_{-24}$~GeV and 60~GeV~$<M_H < 1000$~GeV are shown in
Table~\ref{stab} and in Figure~\ref{sq}.  The agreement is
remarkable.
\begin{figure}
\small \def\baselinestretch{1} \normalsize
\postscript{xxxq.ps}{0.7}
\caption[]{$\sin^2\hat{\theta}_W(M_Z)$ obtained from various observables
assuming $m_t = 150^{+19}_{-24}$~GeV, 60~$< M_H < 1000$~GeV.}
\label{sq}
\end{figure}

One can also extract the radiative correction parameter
$\Delta r$ (eqn.  (\ref{mwz})).  One finds
\beq \Delta r = 0.049 \pm 0.008 \label{eq55} \eeq
compared to the expectation $0.0626 \pm 0.0009 (0.0273)$ for $m_t =
100(200)$, $M_H = 300$.   Similarly, in the
$\overline{\rm MS}$ scheme, one finds
\beq    \Delta \hat{r}_W = 0.069 \pm 0.006, \eeq
compared with the expectation $0.0696 \pm 0.0009 (0.0723)$.

The hadronic $Z$ width depends on the value of \alsz. The quoted results
use the value $0.12 \pm 0.01$ obtained from $Z$-decay event topologies
and low energy data \cite{v9}. One can also obtain a value of \alsz from the
hadronic widths, and in particular from $R$, which is insensitive to $m_t$. A
fit to all $Z$-pole and other data (but not including the constraint \alsz =
$0.12 \pm 0.01$) to
\alsz, \siz, and $m_t$ yields \alsz = $0.130 \pm 0.009$, which is consistent
but slightly above the other determinations.  From
$M_Z, \Gamma_Z, R, \sigma^h_p, $ and $\Gamma_{b \bar{b}}$ only,
one finds the higher value $0.135 \pm 0.011$.
When \alsz = $0.12 \pm 0.01$ is
included as a separate constraint in the fit to all data
one obtains the average  $\alsz = 0.126 \pm 0.007$.
These values are listed in Table \ref{als}, along with
the most important low energy determinations \cite{v9}. There is a
slight tendency for higher values from the $Z$-pole data, but given
the uncertainties (which are usually dominated by
theoretical errors) there is no real discrepancy.
\begin{table}\centering \small \def\baselinestretch{1} \normalsize
\begin{tabular}{|cc|} \hline
\alsz  & source \\
\hline
$0.130 \pm 0.009$     & precision $Z$-pole and low energy \\
$0.135 \pm 0.011$     & $M_Z, \Gamma_Z, R, \sigma^h_p, \Gamma_{b \bar{b}}$
 \\ \hline \hline
$0.123 \pm 0.005$     & event topologies \cite{v9} \\
$0.118 \pm 0.005$      &  $\tau$ decays \cite{v9} \\
$0.112 \pm 0.005$      &  deep inelastic scattering (DIS) \cite{v9} \\
$0.113 \pm 0.006$      &  $\Upsilon,\ J/\psi$ \cite{v9} \\ \hline
$0.12 \pm 0.01$      &  event topologies, $\tau$, DIS,
   $\Upsilon,\ J/\psi$ \\ \hline  \hline
$ 0.126 \pm 0.007$ & combined \\
\hline
\end{tabular}
\caption[]{Values of \alsz from indirect precision data,
event topologies, low energy data, and all data.}
\label{als}
\end{table}

The value of \alsz \ from the precision experiments is strongly
anticorrelated with \mt, as can be seen in Figure \ref{alsmt}.
In particular, larger \mt \ corresponds to smaller \alsz, in
better agreement with the low energy data.
\begin{figure}
\small \def\baselinestretch{1} \normalsize
\postscript{xxalsmt.ps}{0.55}
\caption[]{90\% CL allowed region in \alsz and \mt \ from a combined
fit to precision $Z$-pole and other data (but not including event
topology and low energy determinations of \alsz).}
\label{alsmt}
\end{figure}

\section{Implications for Grand Unification}
These results are in excellent agreement with the predictions
of grand unification in the minimal supersymmetric extension of the
(MSSM), but not in the simplest (and most predictive)
non-supersymmetric GUT (SM) \cite{recent}.
In particular, using
$\alpha^{-1}(M_Z) = 127.9 \pm 0.2$ and \alsz = 0.12 $\pm$ 0.01
one predicts
\beqa \sin^2\hat{\theta}_W (M_Z) & = & 0.2334\pm
0.0025 \pm 0.0025\ {\rm (MSSM)}, \nonumber \\
\sin^2\hat{\theta}_W (M_Z) & = & 0.2100\pm 0.0025 \pm
0.0007 \ {\rm (SM)}, \label{eqgut} \eeqa
where the first uncertaintly is from $\alpha_s$ and $\alpha^{-1}$,
and the second is an estimate of theoretical uncertainties
from the superspectrum, high-scale thresholds, and possible
non-renormalizable operators \cite{ipol}.
The MSSM prediction is in excellent agreement with the
experimental value $0.2326 \pm 0.0006$, while the SM prediction is
in conflict with the data.
Because of the large uncertainty in \alsz, it is convenient
to invert the logic and use the precisely known
$\alpha^{-1}$ and $\sin^2\hat{\theta}_W (M_Z)$
to predict \alsz:
\beqa \alsz & = & 0.125 \pm 0.002 \pm 0.009 \ {\rm (MSSM)}, \nonumber \\
\alsz & = & 0.072 \pm 0.001 \pm 0.001 \ {\rm (SM)}, \label{eqgut1} \eeqa
where again the second error is theoretical.
 It is seen that the
 SUSY case is in excellent agreement with the experimental
 \alsz = 0.12 $\pm$ 0.01, while the
 simplest ordinary GUTs are excluded
 (this is completely independent of proton decay).
 The unification slightly prefers larger values of \alsz, as suggested
 by the $Z$-pole data, but the theoretical uncertainties are
 comparable to the error on the observed \alsz (which is also
 dominated by theory). Proton decay is strongly
 suppressed in SUSY-GUTs. Perhaps, the coupling constants will
 indeed prove to be the ``first harbinger of supersymmetry''
 \cite{amaldi}.

\section{Conclusions}
 \begin{itemize}
 \item There is no evidence for any deviation from the standard model.
 \item   \msb: $\siz = 0.2328 \pm 0.0007$
\item On-shell: $\sinn \equiv 1 - \mw^2/\mz^2 = 0.2267 \pm 0.0024$,
where the uncertainties are mainly from \mt.
\item In the standard model one predicts:
$\mt = 150^{+19 + 15}_{-24 - 20}$ GeV, where the central value assumes
\mh = 300 GeV and the second uncertainty is for \mh $\ra$ 60 GeV ($-$) or
1 TeV (+).
\item In the
MSSM $\mt = 134^{+23}_{-28} \pm 5$ GeV, where the difference
is due the light Higgs scalar expected in the MSSM.
\item Precision data yield the 95\% CL constraints
\beq 83 \ {\rm GeV} < \mt < 197 \ {\rm GeV}, \eeq
where the lower (upper) limits are for \mh \ = 60 (1000) GeV.
The lower limit is valid for any decay mode, and is to be compared with the
direct CDF limit $\mt > 91$ GeV, which assumes canonical decays.
\item There is no significant
constraint on \mh \ until \mt \ is known independently.
\item  Precision $Z$-pole and low-energy data yield the indirect
result $\alsz = 0.130 \pm 0.009$, in reasonable agreement with the
value $0.12 \pm 0.01$ obtained from jet event topologies and low
energy direct determinations.
\item
The low energy couplings are in excellent agreement with
the predictions of supersymmetric grand unification, but not with
the simplest (and most predictive) non-supersymmetric grand unified
theories.
\item The precision data place stringent limits on many types of new
physics into the TeV range.
\item In the future, precision electroweak experiments will be a useful
complement to high energy colliders.
 \end{itemize}

\end{document}